\documentclass[twocolumn,aps]{revtex4}
\usepackage{bm}
\begin{document}
\title{Low temperature decoherence by
electron-electron interactions: \\
Role of quantum fluctuations }
\author{Dmitri S. Golubev$^{1,3}$ and Andrei D. Zaikin$^{2,3}$}
\affiliation{$^1$Institut f\"ur Theoretische Festk\"orperphysik,
Universit\"at Karlsruhe, 76128 Karlsruhe, Germany \\
$^2$Forschungszentrum Karlsruhe, Institut f\"ur Nanotechnologie,
76021 Karlsruhe, Germany\\
$^3$I.E.Tamm Department of Theoretical Physics, P.N.Lebedev
Physics Institute, 119991 Moscow, Russia}

\begin{abstract}
We derive a general expression for the conductivity of a
disordered conductor with electron-electron interactions (treated
within the standard model) and evaluate the weak localization
correction $\delta\sigma_{\rm wl}$ employing no approximations
beyond the accuracy of the definition of $\delta\sigma_{\rm wl}$.
Our analysis applies to all orders in the interaction and extends
our previous calculation by explicitly taking into account quantum
fluctuations around the classical paths for interacting electrons
(pre-exponent). We specifically address the most interesting low
temperature limit and demonstrate that such fluctuations can only
be important in the perturbative regime of short times while they
are practically irrelevant for the Cooperon dynamics at longer
times. We fully confirm our conclusion about the existence of
interaction-induced decoherence of electrons at zero temperature
for the problem in question. We also demonstrate irrelevance of a
perturbative calculation by Aleiner {\it et al.} (AAV) [J. Low
Temp. Phys. {\bf 126}, 1377 (2002)] and refute AAV's critique of
our earlier analysis.

\end{abstract}

\maketitle




\section{Introduction}
The electron decoherence time in
disordered conductors saturates to a finite value at low temperatures
\cite{MW}. In Ref. \cite{GZ1} we offered an explanation of this effect
attributing it to electron-electron interactions. This explanation was
supported by a detailed nonperturbative calculation \cite{GZ2}
and the results \cite{GZ1,GZ2} are in a good agreement with
experimental findings \cite{MW}. Subsequently Aleiner {\it et al.} (AAG)
developed an alternative -- perturbative in the interaction --
calculation \cite{AAG} and claimed that the results of the latter ($i$)
contradict to our results \cite{GZ1,GZ2} and ($ii$) yield zero
electron dephasing
rate at $T \to 0$. AAG also claimed (Sec. 6.1 and 6.2 of
\cite{AAG}) that they have found a ``mistake'' in our
calculation \cite{GZ2}.

All these claims \cite{AAG} have been carefully analyzed and
demonstrated to be in error \cite{GZ3}. We have argued that
(a) the perturbative calculation \cite{AAG} yields ambiguous results
and, hence, is useless for the problem of electron dephasing
at low temperatures, (b) even if one adopts the perturbative
strategy \cite{AAG} one
recovers a {\it finite} electron dephasing time at $T \to 0$, (c)
on a perturbative level our results \cite{GZ2} {\it do agree} with those
of Ref. \cite{AAG} and (d) the claim \cite{AAG} about ``missing
diagrams'' in our calculation is wrong. This discussion and further
arguments on low temperature dephasing due to electron-electron
interactions were reviewed in Ref. \cite{GZS}.

Recently Aleiner {\it et al.} (AAV) \cite{AAV} have made another
attempt to challenge our results and conclusions. One of the goals
of the present paper is to analyze and refute these new criticisms.
In particular, we will point out that the replacement of
``the density matrix by its Wigner transform'', eq. (5) of \cite{AAV}, 
claimed by AAV as ``the main source of the mistake'' {\it was not 
performed at all} in our derivation of the effective action, cf. eq. (43) of
Ref. \cite{GZ2}. We will also analyze and prove irrelevant another
suggestion of AAV, that the term $S_{\rm R}$ in the effective action
\cite{GZ2} should contain an imaginary part which -- according to AAV
-- would be responsible for the well-known cancellation of 
``coth'' and ``tanh'' in the first order perturbation theory at $T \to 0$. 

Since AAV's critique
of our calculation is essentially restricted to these two
claims \cite{FN0,sec}, the above observations are already sufficient
to conclude this discussion. However, taking into
account fundamental importance of the issue, we have performed
an additional analysis aimed to construct a complete solution of the
problem. Our main goal is to evaluate the weak localization (WL)
correction to the conductivity in the presence of interactions making
{\it no approximations} beyond the accuracy of the {\it definition}
of this quantity. This solution is worked out below and our main result
is presented in eqs. (\ref{sigmafinal})-(\ref{me}).

Our paper is organized as follows. After brief remarks on physics
and experiment (Section 2) we demonstrate (in Section 3) that AAV's
perturbative calculation \cite{AAV} is unsuitable for
the problem of quantum decoherence of electrons at low $T$.
In Section 4 we briefly recollect our earlier results and,
making use of general arguments, explain why our path integral
analysis \cite{GZ2,GZ3,GZS} is sufficient for the problem in
question. The main results of this paper are presented in Section 5.
We first derive a formally exact expression for the conductivity of an arbitrary
disordered conductor in the presence of electron-electron interactions
(Sec. 5A) and then use it to explicitly evaluate the WL correction
to all orders in the interaction (Sec. 5B and 5C). This
analysis confirms our previous results \cite{GZ2,GZ3} and extends
them by fully accounting for quantum fluctuations around the classical
paths for interacting electrons. We compare our results with
those of other authors and present further discussion in Sec. 5D and 5E.
In Section 6 we specifically address and refute AAV's
critique \cite{AAV} of our earlier calculation. A brief summary is
presented in Section 7. Some technical details are relegated to Appendix.

\section{Physics and experiment}
In Ref. \cite{AAV} AAV pointed out that our
results and conclusions \cite{GZ1,GZ2,GZ3,GZS} are ``physically
inconsistent'' with the qualitative arguments \cite{AAG} against quantum
dephasing in the zero temperature limit. According to these authors 
quantum decoherence would be impossible in a subsystem
of any interacting quantum system provided the latter is close
to equilibrium at $T \to 0$. At this point we note that the
arguments \cite{AAG} contradict not only to our conclusions but also 
to numerous results for various other models -- including exactly
solvable ones -- where quantum decoherence of one degree of freedom is
obtained as a result of its interaction with others even at $T \to 0$,
see, e.g., \cite{Weiss}. Hence, no general proof can be 
constructed which would rule out
interaction-induced quantum decoherence at $T \to 0$.  
The conclusion about the presence or absence of zero temperature
dephasing in any given model can only be obtained from a
detailed calculation, not on the basis of ``general arguments''.

AAV also claimed \cite{AAV} that there exists an ``overwhelming
experimental evidence'' against our statements. By means of a detailed
comparison with the experimental data we have demonstrated that our
predictions are in a good quantitative agreement with numerous
experiments (see, e.g., Refs. \cite{GZ1,GZ2,GZS,GZ98}) {\it including} those
(see Section 5 of Ref. \cite{GZS}) where for some samples
no saturation of $\tau_{\varphi}$ was claimed down to $T \sim 50$ mK.
Therefore, a bulk piece of existing experimental results
-- including very recent ones \cite{Ba,Mo} -- clearly supports the
conclusion about the presence of interaction-induced
dephasing at low $T$ rather than argues against it.

\section{Insufficiency of AAV's perturbative approach}

In Refs. \cite{GZ3,GZS} we have already explained in details why
perturbative in the interaction techniques -- especially if combined
with the Golden rule approximation -- are insufficient
for the problem in question. However, since AAV still keep using
such techniques, we will briefly repeat our arguments adopting
them to the calculation \cite{AAV}.

In Ref. \cite{AAV}
AAV have related the dephasing time $\tau_{\varphi}$ for an electron
in a disordered conductor to the self-energy for the Cooperon.
Expressing the Cooperon in the form
\begin{equation}
{\cal C}(\omega,Q,\epsilon)=
\frac{1}{-i\omega + DQ^2 +\Sigma(\omega,Q,\epsilon)},
\label{1}
\end{equation}
they have defined the dephasing time at $T=0$ as follows
\begin{equation}
\tau_{\varphi}^{-1}=\Sigma (\omega=0,Q=0,\epsilon),
\label{sigma}
\end{equation}
see the unnumbered equation after eq. (2) of \cite{AAV}. This
definition is ambiguous since the
self-energy is a function of $\epsilon$ whereas $\tau_{\varphi}$ in
our problem is a function of temperature $T$ but not of
$\epsilon$. The ambiguity disappears only at $T \to 0$, since according to
Eq. (1) of Ref. \cite{AAV} in this limit one should set $\epsilon \to 0$.
Another drawback of the definition (\ref{sigma}) is that 
the self-energy is evaluated only in the limit $Q=0$. 
However, the dependence of $\Sigma$ on the momentum $Q$
in eq. (\ref{1}) cannot be {\it a priori} neglected.   

In addition to the above approximations, AAV suggested to
replace the exact self-energy in Eq. (\ref{sigma})
by the result of the first order perturbation theory in the interaction
$\Sigma^{(1)}(\omega,Q,\epsilon):$
\begin{equation}
\frac{1}{\tau_\varphi^{\rm AAV}}=\Sigma^{(1)}(\omega=0,Q=0,\epsilon=0).
\label{01}
\end{equation}
This suggestion constitutes their {\it major mistake}. Quite obviously, 
it is not possible to recover the unknown function 
(in our case $\Sigma$) if one only evaluates the first order 
term ($\Sigma^{(1)}$) of its Taylor expansion.

In order to
illustrate this point it suffices to consider the following simple example.
Assume, for instance, that the ``Cooperon'' $\tilde {\cal C}(t)$
depends weakly on the coordinates and neglect this dependence.
We also assume that
in the presence of interactions this ``Cooperon'' decays in time as
\begin{equation}
\tilde {\cal C}(t)=\theta(t)(1+\alpha t){\rm e}^{-(\alpha+\beta T)t},
\label{C1}
\end{equation}
where $\alpha$ and $\beta$ are proportional to the interaction strength, i.e
(\ref{C1}) reduces to $\theta(t)$ in the absence of interactions.
After the Fourier transformation of (\ref{C1}) one readily finds
\begin{eqnarray}
\tilde {\cal C}(\omega)&=&\frac{1}{-i\omega +\tilde\Sigma(\omega)},
\nonumber\\
\tilde\Sigma(\omega)&=&\frac{(\alpha +\beta T)^2 -i\omega \beta
T}{2\alpha +\beta T -i\omega }.
\label{nonpert}
\end{eqnarray}
Combining (\ref{nonpert}) with the definition (\ref{sigma}) we obtain
\begin{equation}
\frac{1}{\tau_\varphi}=\frac{\alpha+\beta T}{1+\frac{\alpha}{\alpha +\beta T}},
\label{real}
\end{equation}
i.e. at $T\to 0$ one arrives at a non-zero dephasing rate
\begin{equation}
\tau_{\varphi}^{-1}=\alpha /2.
\label{r0}
\end{equation}

Let us now evaluate $\tau_\varphi$ for the same example following the approach of
AAV. For this purpose we expand the exact expression for $\tilde\Sigma(\omega)$
(\ref{nonpert}) in powers of interaction and get
\begin{equation}
\tilde\Sigma(\omega)=\beta T+\frac{\alpha^2}{-i\omega}+\dots.
\label{pert2}
\end{equation}
Keeping only the first order contribution to the self-energy, from AAV's eq.
(\ref{01}) we find
\begin{equation}
\frac{1}{\tau_{\varphi}^{\rm AAV}}=\beta T.
\label{AAV}
\end{equation}
This result differs drastically from the exact one (\ref{real})
at sufficiently low temperatures. In particular, at $T \to 0$ the
dephasing rate (\ref{AAV}) vanishes, while the exact expression
approaches a non-zero (linear in the interaction) value (\ref{r0}).
Furthermore, all higher order terms in the expansion (\ref{pert2})
do not vanish at $T \to 0$ and, moreover, diverge at small $\omega$.

The reason for the failure of AAV's perturbative approach is, of course,
obvious from eq. (\ref{nonpert}): An expansion of this expression in
powers of $\alpha$ and $\beta$ is only justified for $\omega \gg
\alpha +\beta T$, i.e. in the limit of high frequencies or short times.
We, in contrast, are interested in the opposite limit of small frequencies
or long times. Thus, AAV simply missed the low temperature contribution to
the dephasing rate by inadequately extending their perturbative expansion of
$\Sigma(\omega,Q,\epsilon)$ to low frequencies, whereas it can be applied
at high frequencies only.

Our simple example, eq. (\ref{nonpert}), also illustrates irrelevance of the
claim \cite{AAV} that our approach ``is equivalent to calculating only a single
contribution $\Sigma^{({\rm b})}$ to the self-energy and using the conventional
Dyson equation''. AAV arrived at this conclusion simply by observing
the combination $\Sigma^{({\rm b})}-\Sigma^{({\rm c-f})} \sim \beta T$  in
their first order perturbative result (cf. eqs. (3,4) in \cite{AAV})
and comparing it with our result,
which contains not only $\beta T$ but also the $T$-independent
contribution. The only -- superficial --
reason for Aleiner {\it et al.} to perform this comparison and to qualify
our results as ``purely perturbative'' is that our expression
for the dephasing rate
``is proportional to the first power of the fluctuation
propagator'' \cite{AAG}. We hope it should be sufficiently clear from eq.
(\ref{nonpert}) that {\it linear} in the interaction
expression (\ref{r0}) is {\it non-perturbative} and, hence,  it cannot
be obtained by a simple expansion (\ref{pert2}) of the
self-energy $\tilde\Sigma (\omega )$ in the interaction.

In order to avoid misunderstandings \cite{FNAAV} let us emphasize that the
above example (\ref{C1}) is {\it not} meant to be an explicit
solution for the
problem with electron-electron interactions. This solution will be
worked out below in Sec. 5. Eq. (\ref{C1}) is just an
illustration of one of the drawbacks of the perturbative approach \cite{AAV}
to the problem of quantum dephasing. In the problem with disorder 
and electron-electron interactions the situation turns out to
be by far more complicated. For instance,
already the first order result diverges both with time and at large
frequencies in 1d and 2d systems, see, e.g., eq. (70) of
Ref. \cite{GZ3}.

\section{Path integral analysis: exponent}
An important advantage of our path
integral approach is the possibility to describe
the long-time behavior of the Cooperon with exponential accuracy,
which is sufficient for the problem in question.
This approach is free from ambiguities inherent to the perturbation
theory  \cite{AAG,AAV}.

We define the kernel of the evolution
on the Keldysh contour in terms of the path integral \cite{GZ2,GZ3,GZS}
\begin{eqnarray}
&&\int{\cal D}{p}_1\int{\cal D}{x}_1
\int {\cal D}{p}_2\int{\cal D}{x}_2\,
\nonumber\\
&\times&
{\rm e}^{\frac{1}{\hbar}(iS_0[{p}_1,{x}_1]
-iS_0[{p}_2,{x}_2]
-iS_{\rm R}[{p}_1,{x}_1,{p}_2,{x}_2]
-S_{\rm I}[{x}_1,{x}_2])}
\label{J}
\end{eqnarray}
Here
$
S_0[{p},{x}]=\int_{0}^{t}dt'
[{p}\dot{{x}}-{{p}^2}/{2m}-U_{\rm imp}({x})]$
is the action of a noninteracting electron in a disordered potential
of impurities and the terms $iS_{\rm R}$ and $S_{\rm I}$ describe the effect of
the bath (formed by all the electrons) on the motion of a single electron. This
form of the effective action for an interacting particle is standard
in the Feynman-Vernon theory of influence functionals
\cite{FH}. The path integral (\ref{J}) is evaluated within a
semiclassical approximation controlled by the parameter $k_Fl \gg 1$.
For a $d$-dimensional conductor we find
\begin{equation}
{\cal C}(t,x=0)=\theta(t)\frac{A_d(t)}{(4\pi Dt)^{d/2}}{\rm e}^{-f_d(t)},
\label{fd}
\end{equation}
where $D$ is the diffusion coefficient, $A_d(t)$ is pre-exponent ($A_d(t)\equiv 1$ without interactions), and the function $f_d(t)$ is obtained by
evaluating the influence functional
on pairs of classical time-reversed paths. The result can be written
in the form \cite{GZ3}
\begin{equation}
f_d(t)=\alpha t +\delta f_d (T,t),
\label{dft}
\end{equation}
where the function $\delta f_d (T,t)$ is $\propto Tt^{3/2}$ for $d=1$ and
$\propto Tt\ln Tt$ for $d=2$ in the limit $Tt \gg 1$, while it is
$\propto \sqrt{t}\ln t$ for $d=1$ and $\propto \ln t$ in the opposite
limit $Tt \ll 1$.
The linear dependence
$\delta f_d (T,t)\propto Tt$ strictly applies for $d=3$.

It is also instructive to explicitly indicate the dependence
of our results on the Plank's constant $\hbar$. At $T\to 0$ we have
\begin{equation}
\ln [{\cal C}_d(t)/{\cal C}_d^{(0)}(t)] =-\frac{S^{\rm (cl)}(t)}{\hbar}+
\ln [A_d(\hbar ,t)],
\label{Scl}
\end{equation}
where $S^{\rm (cl)}\simeq at$ is the classical ($\hbar$-independent)
action on time-reversed saddle point paths ($a$ is $\hbar$-independent
and $\alpha \equiv a/\hbar$) and $\hbar \ln A_d$
represents the quantum correction to the classical action. This
quantum correction can
only be important if $S^{\rm (cl)}$ is small as compared to $\hbar$,
i.e. at times $t \ll \hbar /a$. The perturbative
approach \cite{AAG,AAV} applies {\it only} in this limit.
On the other hand, for nonzero $t$ the quantum correction $\hbar \ln
A_d$ should vanish at $\hbar \to 0$. Already because of
this reason it cannot cancel the classical part of the
action $S^{\rm (cl)} \gtrsim \hbar$.
It is therefore sufficient to evaluate the $\hbar$-independent part
of $-iS_{\rm R}-S_{\rm I}$ on pairs of classical time-reversed paths
and obtain the dephasing time from the condition $S^{\rm
(cl)}(\tau_{\varphi})\sim \hbar$. Furthermore, it turns out that
$S_{\rm R}$ vanishes for such paths \cite{GZ2} implying
that $S_{\rm R}$ can only contribute to the {\it
pre-exponent} but not to $S^{\rm (cl)}$. Hence, $S_{\rm R}$ is
irrelevant for $\tau_{\varphi}$ and the idea
that quantum fluctuations around the time-reversed classical
paths generate the $S_{\rm R}$-dependent contribution to
the classical action can be rejected
on general grounds without any calculation \cite{Fl}.

This is the logics behind our saddle point analysis \cite{GZ2,GZ3} which
AAV \cite{AAV} attempted to challenge. The only way to support the AAV's
arguments is to prove that at $T \to 0$ the term $S_{\rm R}$ provides
a contribution to $A_d$ proportional to $\exp (at/\hbar)$
which grows {\it exponentially} with time and {\it diverges} in
the classical limit $\hbar \to 0$ [in Ref. \cite{AAV} such a
contribution was claimed to be provided by the terms
$\Sigma^{({\rm c-f})}$].
Below we will explicitly evaluate not only $S^{\rm (cl)}$ but also
the quantum correction to the effective action in all orders in the
interaction. We will demonstrate that --
in contrast to the AAV's claims -- the  $S_{\rm R}$-dependent
pre-exponent cannot grow at sufficiently long times and, hence,
in no way can compensate an exponentially decaying contribution
from the $S_{\rm I}$-terms.

\section{Decoherence by interactions}

\subsection{Exact results}
In the beginning we will closely follow the analysis
of Ref. \cite{GZ2} where the interested reader can find further
details. We start from the general quantum mechanical expression for the linear
conductivity $\sigma$ which can be written in the form
\begin{equation}
\sigma=\frac{e}{3i\hbar}\int\limits_{-\infty}^t dt'
\left\langle{\rm tr}\left(\hat{    j}( {x})
\hat U_1(t,t')[\hat {   x},\hat \rho_V(t')]\hat U_2(t',t)
\right)\right\rangle_{V}
\label{sigma1}
\end{equation}
Here the current density operator is defined via
\begin{equation}
\langle {x}_1|\hat{{   j}}({x})|{x}_2\rangle
=\frac{\hbar e}{im}\left[\nabla_{{x}_1}\delta_{{x}_1,{x}}
\delta_{{x}_2,{x}}-
\delta_{{x}_1,{x}}\nabla_{{x}_2}\delta_{{x}_2,{x}}\right]
\label{tok}
\end{equation}
and $\langle ... \rangle_V$ implies averaging over the fluctuating
quantum fields $V^+$ and $V^-$
which mediate Coulomb interaction between electrons.
In eq. (\ref{sigma1}) and below we implicitly assume averaging
over the coordinate $x$ where the current density is evaluated.

The evolution operators $\hat U_{1,2}$ in (\ref{sigma1}) are
\begin{equation}
\hat U_{1,2}(t,t')=\bm{T}
\exp\left[-\frac{i}{\hbar}\int_{t'}^{t}d\tau\, \hat H_{1,2}(\tau )
\right],
\label{U12}
\end{equation}
where
\begin{eqnarray}
\hat H_1(t)=\hat H_0-\mu-e\hat V^+(t) -\frac{1}{2}[1-2\hat\rho_V(t)]e\hat V^-(t),
\nonumber\\
\hat H_2(t)=\hat H_0-\mu-e\hat V^+(t) +\frac{1}{2}e\hat V^-(t)[1-2\hat\rho_V(t)],
\label{H12}
\end{eqnarray}
and $\hat H_0=\hat p^2/2m + U_{\rm imp}(\hat x)$ is the Hamiltonian
of a non-interacting electron.
The density matrix $\rho_V(t')$ obeys the non-linear equation
\begin{eqnarray}
i\hbar\frac{\partial\hat\rho_V}{\partial t}&=&[\hat H_0-e\hat V^+,\hat\rho_V]
-\frac{1}{2}(1-\hat\rho_V)e\hat V^-\hat\rho_V
\nonumber\\
&&-\,\frac{1}{2}\hat\rho_V e\hat V^-(1-\hat\rho_V).
\label{Liouville}
\end{eqnarray}

Averaging of $\rho_V(t)$ over $V^{\pm}$ yields the
exact single electron density matrix in the presence of disorder and
Coulomb interactions. The next steps in Refs. \cite{GZ2,GZS} were to
express the kernels of the operators $\hat U_{1,2}$ in terms of
the path integrals and average their product over the fluctuating
fields $V^{\pm}$. After that one arrives at the path integral (\ref{J}).

At this point we depart from the analysis of Refs. \cite{GZ2,GZS}.
We will postpone using the path integrals and continue exact manipulations
with the operators. We first note that the solution of eq. (\ref{Liouville})
with the initial
condition $\hat\rho_V(0)=\hat\rho_0\equiv [1+{\rm e}^{(\hat H_0-\mu)/T}]^{-1}$ can be expressed in the following
exact form
\begin{equation}
\hat \rho_V(t)=\left[1+\hat u_2(t,0){\rm e}^{(\hat H_0-\mu)/T}\hat u_1(0,t)\right]^{-1},
\label{rV}
\end{equation}
where we have defined
\begin{equation}
\hat u_{1,2}(t,t')=\bm{T}\exp\left[-\frac{i}{\hbar}\int_{t'}^{t}d\tau\,
\hat{\tilde H}_{1,2}(\tau ) \right]
\label{u12}
\end{equation}
and
\begin{equation}
\hat{\tilde H}_{1,2}=\hat H_{0}(t')-\mu-e\hat V^+(t')\mp e\hat V^-(t')/2.
\label{ham}
\end{equation}
We then observe that the operators $\hat U_{1,2}$ satisfy the
Schr\"odinger equation
\begin{eqnarray}
i\hbar\frac{\partial}{\partial t}\hat U_1(t,t')=\hat H_1(t)\hat U_1(t,t'), \;\; \hat U_1(t',t')=\hat 1,
\nonumber\\
i\hbar\frac{\partial}{\partial t}\hat U_2(t',t)=\hat U_2(t',t)\hat H_2(t), \;\; \hat U_2(t',t')=\hat 1.
\label{Schr}
\end{eqnarray}
The solutions of (\ref{Schr}) can be found exactly. They are
\begin{eqnarray}
\hat U_1(t,t')=[1-\hat\rho_V(t)]u_1(t,t')[1-\hat\rho_V(t')]^{-1},
\nonumber\\
\hat U_2(t',t)=[1-\hat\rho_V(t')]^{-1}u_2(t',t)[1-\hat\rho_V(t)].
\label{ex1}
\end{eqnarray}
Combining these expressions with eq. (\ref{rV}) one can rewrite the
operators (\ref{ex1}) in the following identical form
\begin{eqnarray}
\hat U_1(t,t')=[(1-\hat\rho_V(t'))\hat u_1(t',t)+\hat \rho_V(t')\hat u_2(t',t)]^{-1},
\nonumber\\
\hat U_2(t',t)=[\hat u_2(t,t')(1-\hat\rho_V(t'))+\hat u_1(t,t')\hat \rho_V(t')]^{-1}.
\label{exact}
\end{eqnarray}
Eqs. (\ref{exact}) is our key technical result. Together with
eqs. (\ref{sigma1},\ref{tok}) this result provides a formally
exact expression for the linear conductivity of an arbitrary
disordered conductor in the presence of electron-electron
interactions. {\it All} the diagrams of the
perturbation theory in {\it all} orders in the interaction are
fully contained in the above expressions and can be
recovered by expanding (\ref{exact}) in $V^{\pm}$ with
subsequent averaging over these fields. For instance, expanding
(\ref{exact}) to the second order in $V^{\pm}$, after averaging
one arrives at contributions of the type $\langle V^+V^+
\rangle$ and  $\langle V^+(1-2\hat\rho )V^-\rangle$ which yield
respectively ``coth'' and ``tanh'' terms in the perturbation
theory \cite{AAG,GZ3}.

Thus, making no approximations we have demonstrated that the
time evolution of the single electron density matrix in the
presence of interactions is determined by the
operators $\hat u_{1,2}(t,t')$ (\ref{u12}) which do not contain
the density matrix $\hat \rho_V$ at all. We also note that the
operators $\hat U_{1,2}$ (\ref{exact}) depend on the density matrix
$\hat \rho_V$ taken at the initial time $t'$ only
but not at later time moments. Below we will
make use of these features and evaluate the WL
correction to conductivity to all orders in the interaction.

\subsection{Quasiclassics}

Let us first prepare the main building blocks of our calculation
of the WL correction. This calculation will then be completed in
Sec. 5C.

To begin with, we notice that in eqs. (\ref{exact}) relative contributions
of the evolution operators  $\hat u_{1,2}(t,t')$ are ``weighted'' by
the factors $1-\hat\rho_V(t')$ and $\hat\rho_V(t')$. In the low
temperature limit almost at any
electron energy one of these factors dominates over the other
\cite{FN1}. Hence, one of the two operators $\hat u_{1}$ or
$\hat u_{2}$ in (\ref{exact}) can be neglected except if the
eigenvalues of $\hat\rho_V(t')$ are close to 1/2. Consider, e.g.,
small eigenvalues of this operator. This situation
describes electrons with energies above the Fermi level. In this case
terms containing $\hat \rho_V(t')$ in (\ref{exact}) can be
neglected and we arrive at the following contribution to
$\langle x_1|\hat U_1(t,t')[\hat x,\hat \rho_V(t')]\hat U_2(t',t)|x_2\rangle$:
\begin{eqnarray}
\int \int dz_1dz_2 \langle x_1| \hat u_{1}(t,t') | z_1\rangle \langle z_2| \hat u_{2}(t',t) | x_2\rangle
\nonumber\\
\times \langle z_1|[\hat x, \hat \rho_V(t')]|z_2\rangle .
\label{dr}
\end{eqnarray}
Let us express the matrix elements of the operators $\hat u_{1,2}$
via the path integrals
\begin{equation}
\langle y_{1,2}| \hat u_{1,2}(t,t') | z_{1,2}\rangle =
\int\limits_{y_{1,2}}^{z_{1,2}}{\cal D} {x (\tau )}
e^{\frac{i}{\hbar}S_{1,2}[x(\tau )]} \; ,
\label{pi}
\end{equation}
where $S_{1,2}$ are the exact actions pertaining to the
Hamiltonians (\ref{ham})
\begin{eqnarray}
S_{1,2}=S_0+e\int_{t'}^{t} d\tau [V^+(\tau,x(\tau))
\pm V^-(\tau,x(\tau))/2],
\label{actions}\\
S_0=\int_{t'}^{t} d\tau\left(\frac{m\dot x^2(\tau)}{2}-U_{\rm imp}(x(\tau))\right).
\label{action0}
\end{eqnarray}
We emphasize again that eqs. (\ref{pi}-\ref{action0}) are exact
and they do not contain the electron density matrix  $\hat\rho_V $ at
all. We should now (a) evaluate the path integrals
(\ref{pi}) and then (b) average the combination (\ref{dr}) over the
fluctuating fields $V^\pm$.

{\it Evaluation of the matrix elements}. Let us make use of the fact that
the WL correction to conductivity is
defined within the accuracy $k_Fl \gg 1$. This inequality is usually
well satisfied in disordered metallic conductors.
Hence, we can evaluate the matrix elements
(\ref{pi}) quasiclassically. Since the actions (\ref{actions}) do not
depend on $\rho_V$ one can conveniently employ a regular expansion
of $S_{1,2}$ in powers of $\hbar$. The path integrals are then
easily evaluated and we arrive at the well-known van Vleck formula
\begin{eqnarray}
\langle x_{1,2}| \hat u_{1,2}(t,t') | z_{1,2}\rangle \hspace{3cm}
\nonumber \\
=\sum_n\sqrt{\left(\frac{i}{2\pi\hbar} \right)^3{\rm det}
\frac{\partial^2 S^{(n)}_{1,2}}{\partial x_{1,2}\partial z_{1,2}}}
\;{\rm e}^{\frac{i}{\hbar} S^{(n)}_{1,2}},
\label{vV}
\end{eqnarray}
where $S^{(n)}_{1(2)}\equiv S^{(n)}_{1(2)}(t,t',x_{1(2)},z_{1(2)})
=S_{1(2)}[\tilde x_{1n(2n)}]$
and $\tilde x_{1,2}$ are the exact least action paths obeying the equations
\begin{equation}
\delta S_{1(2)}[\tilde x_{1n(2n)},V^{\pm}]/\delta \tilde x_{1n(2n)}=0
\label{Newton}
\end{equation}
with the boundary conditions  $\tilde x_{1n(2n)}(t')=x_{1(2)}$ and
$\tilde x_{1n(2n)}(t)=z_{1(2)}$. In general there exist several or even many
different classical paths satisfying the above conditions. Here and below
the index $n$ labels all such paths.

Clearly, eq. (\ref{vV}) accounts not only
for the saddle point trajectories (exponent) but also for quantum fluctuations
around $\tilde x_{1n,2n}(\tau )$ (pre-exponent) for arbitrary
$V^{\pm}$. It is also completely obvious that there is no way how
the pre-exponent can cancel the exponent for any configuration
of the fields $V^{\pm}$. Hence, such cancellation is impossible
also after averaging over these fields no matter what the details
of this averaging procedure are.

{\it Averaging over the fluctuating fields}. In general averaging of
the combination (\ref{dr}) over $V^\pm$ involves path integrals
(\ref{av}) over these fields at all times from zero to $t$. This is
because $\hat \rho_V(t')$ is nonlocal in time: According to eq.
(\ref{rV}) it depends on times between zero and $t'$. However,
with the accuracy $k_Fl \gg 1$
one can perform averaging in (\ref{dr}) at times smaller and larger
than $t'$ separately. This splitting is achieved by expressing
$\rho_V(t')$ (\ref{rV}) via the path integrals, making use of
eqs. (\ref{pi}-\ref{action0}) and averaging the whole combination
(\ref{dr}) over $V^{\pm}$ at all times between 0 and $t$. One
arrives at the effective actions containing nonlocal in time contributions
$S_{\rm R,I}(t_1,t_2)$ which vanish for all the trajectories relevant
in the quasiclassical limit $k_Fl \gg 1$ provided $t_1 >t'$, $t_2<t'$
and vice versa. As a result we obtain
\begin{eqnarray}
\int \int dz_1dz_2 J_{12} (t,t';x_{1,2},z_{1,2}) (z_1-z_2) \rho (z_1,z_2) ,
\label{dr2}\\
J_{12}=
  \langle\langle x_1| \hat u_{1}(t,t') | z_1\rangle \langle z_2| \hat
  u_{2}(t',t) | x_2\rangle \rangle_V .
\label{J12}
\end{eqnarray}
Here we used $\langle z_1|[\hat x, \hat \rho ]|z_2\rangle \equiv (z_1-z_2)\rho
(z_1,z_2)$, where $\hat \rho = \langle \hat \rho_V(t')\rangle_V$ is the
exact equilibrium electron density matrix in the presence of
interactions. We can also add that, as it was explained in Sec. 4 of
Ref. \cite{GZS}, with the same accuracy $k_Fl \gg 1$ one can replace
\begin{equation}
\hat \rho_V(t')\longrightarrow\hat \rho_0=[1+{\rm e}^{(\hat
H_0-\mu)/T}]^{-1}
\label{R0}
\end{equation}
already before averaging over $V^\pm$. After that the factorization
(\ref{dr2}) (with $\rho \to \rho_0$) is, of course, an exact procedure.

What remains is to average the product of the two matrix
elements in (\ref{J12}). This averaging is carried out in a standard manner.
If the fields $V^\pm$ vary at scales exceeding the elastic mean free path,
one can neglect the dependence of both  the classical paths $\tilde x_{1,2}$ and
the pre-exponent in eq. (\ref{vV}) on the fields $V^{\pm}$. This approximation
is sufficient for evaluation of the WL correction to the conductivity.

Then averaging with the action (\ref{Sem}) can be performed
exactly. We first integrate over $V^+$. As both actions (\ref{Sem})
and (\ref{actions}) are linear in $V^+$, this integration yields
the $\delta-$function
$$\delta (V^-(\tau,x)-V_0(\tau,x,\tilde x_{1n}(s), \tilde x_{2m}(s)),$$
where
\begin{eqnarray}
V_0(\tau,x,\tilde x_{1n}(s), \tilde x_{2m}(s))=
-e\int\limits_{t'}^t ds[R(s-\tau,\tilde x_{1n}(s)-x)
\nonumber \\
-R(s-\tau,\tilde x_{2m}(s)-x)]\hspace{2cm}
\label{V0}
\end{eqnarray}
and the function $R(t, x)$ is defined in (\ref{+-}).
Due to this $\delta-$function the subsequent integration over
$V^-$ also becomes trivial and we obtain $J_{12}=\sum_{n,m}J_{12}^{nm}$, where
\begin{eqnarray}
J_{12}^{nm}=
\frac{1}{(2\pi\hbar)^3}
\sqrt{{\rm det}\frac{\partial^2 S_{0}^{(n)}}{\partial x_{1}\partial z_{1}}\,
{\rm det}\frac{\partial^2 S_{0}^{(m)}}{\partial x_{2}\partial z_{2}}}
\hspace{1cm}
\nonumber\\ \times
{\rm e}^{\frac{1}{\hbar}(iS_0[\tilde x_{1n}]
-iS_0[\tilde x_{2m}]
-i\tilde S_{\rm R}[\tilde x_{1n},\tilde x_{2m}]
-S_{\rm I}[\tilde x_{1n},\tilde x_{2m}])}.
\label{J2}
\end{eqnarray}
In eq. (\ref{J2}) 
the term $S_{\rm I}$ is identical to one derived in Ref. \cite{GZ2}
(see eq. (55) of that paper) while the action $\tilde S_{\rm R}$ is
obtained from eq. (54) of \cite{GZ2} by formally setting the
function $n( \bm{p},\bm{r})$ equal to zero in that formula. The
action $\tilde S_{\rm R}$ is purely real for any
pair of paths $\tilde x_{1n}$ and $\tilde x_{2m}$. Thus, together with
the terms $iS_0$ it can only provide oscillations of the kernel $J$
and in no way can compensate its decay
$J^{nm}_{12} \propto \exp(-S_{\rm I}/\hbar)$. As it was already discussed in
Ref. \cite{GZ2}
and elsewhere, the action $S_{\rm I}$ is real and positive for {\it any} pair
of trajectories (except for identical ones in which case $S_{\rm I}=0$).
The length of electron trajectories in a metal always grows 
with time since electrons move with a constant velocity $\sim v_F$.
Hence, for any pair
of time-reversed paths $\tilde x_{1n}(s)=\tilde x_{2m}(t+t'-s)$ the
action $S_{\rm I}$ 
grows with time as well. This in turn implies that for such paths
the kernel $J_{12}^{nm}$ decays with time and vanish in the long time 
limit $t-t' \to \infty$ at any temperature including, of course, $T=0$.

In the above analysis we neglected the evolution operator $\hat u_2$
($\hat u_1$) in the exact expression for $\hat U_1$ ($\hat U_2$) (\ref{exact}).
This is correct at low $T$ and for the electron energies above the Fermi level.
Below the Fermi energy, on the contrary, 
one can drop terms containing $1-\hat\rho_V(t')$ because in this case
the eigenvalues of $\hat\rho_V(t')$ are close to one. Then the whole
analysis is repeated, one should only interchange the
operators $\hat u_1$ and $\hat u_2$. In this way one again arrives at
eq. (\ref{J2}) (with $\tilde S_{\rm R} \to
-\tilde S_{\rm R}$) which again decays
as $\exp(-S_{\rm I}/\hbar)$. The remaining options are to neglect either
$\hat u_1$ or $\hat u_2$ in both expressions for $\hat U_{1,2}$
(\ref{exact}). One again finds the contributions $\propto
\exp(-S_{\rm I}/\hbar)$. Since in all these cases $S_{\rm I}$ remains
the same, one concludes that
if the operators $\hat u_1$ and $\hat u_2$ yield comparable
contributions (in which case the exact form of (\ref{exact}) should be
used) they will also decay as $\exp(-S_{\rm I}/\hbar)$ on any pair of
time-reversed paths. Below we present an explicit calculation of the
WL correction which fully confirms this conclusion.

\subsection{Pre-exponent and weak localization correction}

We are now prepared to evaluate the conductivity.
As before, we assume that $k_Fl \gg 1$
and that the fields $V^\pm$ vary in space at scales exceeding
the elastic mean free path $l$. In this case
quasiclassical electron trajectories are not disturbed by
interactions, and the contributions of the fluctuating fields $V^\pm$
add up independently. Therefore, we can approximately split the operators
\begin{equation}
\hat u_{1,2}(t,t')\simeq \hat u_0(t,t')\hat s(t,t', V^+)\hat s(t,t',\pm V^-/2),
\label{split1}
\end{equation}
where $\hat u_0(t,t')$ is the evolution operator pertaining to the
non-interacting Hamiltonian and
\begin{eqnarray}
\hat s(t,t',V)=\hat u_0(t',t)\bm{T}\exp\left[-\frac{i}{\hbar}\int_{t'}^td\tau
(\hat H_0-e\hat V(\tau))\right].
\end{eqnarray}
Within the same accuracy we can replace
\begin{eqnarray}
\hat u_0(t,t')\hat s(t,t', V^+)\simeq \hat u(t,t',V^+)\hspace{2.8cm}
\nonumber\\
=\bm{T}\exp\left(-\frac{i}{\hbar}\int_{t'}^td\tau [\hat H_0-e\hat V^+(\tau)]\right).
\label{repl}
\end{eqnarray}

Combining (\ref{split1})-(\ref{repl}) with (\ref{exact}) we obtain
\begin{eqnarray}
\hat U_1(t,t')&\simeq & \hat u(t,t',V^+)
\big\{(1-\hat\rho_V(t'))\hat s^{-1}(t,t',V^-/2)
\nonumber\\ &&
+\hat \rho_V(t')\hat s^{-1}(t,t',-V^-/2)\big\}^{-1},
\nonumber\\
\hat U_2(t',t)&\simeq &\big\{\hat s(t,t',-V^-/2)(1-\hat\rho_V(t'))
\nonumber\\&&
+\hat s(t,t',V^-/2)\hat \rho_V(t')\big\}^{-1}
 \hat u(t',t,V^+).
\label{rhoVapp}
\end{eqnarray}
Substituting (\ref{rhoVapp}) into eq. (\ref{sigma1}), evaluating the
matrix elements of the operator  $u(t,t',V^+)$ by means of the van
Vleck formula (\ref{vV}) and integrating over the fluctuating fields
$V^{\pm}$ exactly as in Sec. 5B \cite{FN2}, we find
\begin{eqnarray}
\sigma
&=&\frac{e^2}{3 m}
\sum_{n,m}\int_{-\infty}^tdt'\,\int dy_1 dy_2\int dz_1 dz_2
\nonumber\\ &&\times\,
(\nabla_{x_1}-\nabla_{x_2})|_{x_1=x_2} J^{nm}(t,t';x_1,x_2;y_1,y_2)  
\nonumber\\ && \times\,
A_1^{nm}(t,t',y_1,z_1,x_1,x_2)(z_1-z_2) \rho_0(z_1,z_2)
\nonumber\\ &&\times\,
A_2^{nm}(t,t',z_2,y_2,x_1,x_2),
\label{sigmafinal}
\end{eqnarray}
where
\begin{eqnarray}
J^{nm}=
\frac{1}{(2\pi\hbar)^3}
\sqrt{{\rm det}\frac{\partial^2 S_{0}^{(n)}}{\partial x_{1}\partial y_{1}}\,
{\rm det}\frac{\partial^2 S_{0}^{(m)}}{\partial x_{2}\partial y_{2}}}
\hspace{1.9cm}
\nonumber\\ \times\,
\exp\left\{\frac{i}{\hbar}S_{0}[\tilde
x_{1n}]-\frac{i}{\hbar}S_{0}[\tilde x_{2m}]
-\frac{1}{\hbar}S_I[\tilde x_{1n},\tilde x_{2m}]\right\}
\label{Jevol}
\end{eqnarray}
and 
\begin{eqnarray}
A_1^{nm}=\langle y_1|\big\{(1-\hat\rho_0)\hat s^{-1}(t,t',V^-/2)\hspace{2.5cm}
\nonumber\\
+\,\hat \rho_0\hat s^{-1}(t,t',-V^-/2)\big\}^{-1}
|z_1\rangle|_{V^-=V_0(\tau,x,\tilde x_{1n},\tilde x_{2m})} ,
\nonumber\\
A_2^{nm}=
\langle z_2| \big\{\hat s(t,t',-V^-/2)(1-\hat\rho_0)\hspace{2.5cm}
\nonumber\\
+\,\hat s(t,t',V^-/2)\hat \rho_0\big\}^{-1} |y_2\rangle
|_{V^-=V_0(\tau,x,\tilde x_{1n},\tilde x_{2m})}
\label{me}
\end{eqnarray}
As before, the paths $\tilde x_{1n}$ and $\tilde x_{2m}$
satisfy the Newton equation (\ref{Newton}) (with $V^\pm =0$) and the
boundary conditions
$\tilde x_{1n(2m)}(t')=y_{1(2)}$ and $\tilde x_{1n(2m)}(t)=x_{1(2)}$.

Eqs. (\ref{sigmafinal})-(\ref{me}) represent the central result of this paper.
They determine the linear conductivity of an arbitrary
disordered conductor to all orders in the electron-electron interaction.
The above equations  are based on the
exact results (\ref{exact}) and are valid in the quasiclassical limit
$k_Fl \gg 1$. We would like to emphasize that
no quasiclassical approximation for the electron density matrix was employed
during our derivation and no averaging over impurities was performed at all.

Let us briefly analyze eqs. (\ref{sigmafinal})-(\ref{me}). According to
the standard arguments two types of classical paths $\tilde x_{1n}$
and $\tilde x_{2m}$, identical and time-reversed ones,
play an important role in the quasiclassical limit $k_Fl \gg 1$. For a pair of
identical paths $\tilde x_{1n}(s)=\tilde x_{2n}(s)$ the two actions $S_0$
in the exponent (\ref{Jevol}) cancel each other, the term $S_{\rm I}$ vanishes
identically and the matrix elements $A_{1,2}$ reduce to $\delta$-functions
$A_{1,2}^{nm}=\delta (y_{1,2}-z_{1,2})$ because $V_0\equiv 0$ in this case.
In this way we recover the well known property that the diffuson does
not decay in time even in the presence of interactions.

Here we are interested in the quantum correction to conductivity
arising from the time-reversed paths $\tilde x_{1n}(s)=\tilde x_{2m}(t+t'-s)$.
For any pair of such paths the actions $S_0$ cancel again but the interaction
term $S_{\rm I}$ is now positive, it grows with time and yields
(exponential) decay of the quantity $J^{nm}$ (\ref{Jevol}) in the long time
limit. The matrix elements
$A_{1,2}^{nm}$ also depend on the interaction and on time in this case.
It is obvious, however, that $A_{1,2}^{nm}$ {\it cannot grow} at long times
because the function $V_0$ (\ref{V0}) is purely real and, hence, $\hat s(t,t',\pm V_0/2)$
are the unitary operators. The matrix elements of such operators can only
oscillate provided the function $V_0$ changes in time. Hence, no
compensation of decaying $J^{nm} \propto \exp (-S_{\rm I}(t-t')/\hbar )$ can be
expected for sufficiently large $t-t'$, and the whole expression
under the integral over $t'$ in  eq. (\ref{sigmafinal}) decays
exponentially together with $J^{nm}$ for any pair of time-reversed paths.
Obviously, the matrix elements (\ref{me}) also cannot grow if one formally
takes the limit $\hbar \to 0$, while $J^{nm}$ vanishes in this limit.
All that implies that -- in full agreement with our general arguments
(Sec. 4) -- one can indeed obtain the dephasing time from the condition
$S_I(\tau_{\varphi}) \sim \hbar$ which does not depend on the density
matrix $\hat \rho_V$ and identically reproduces our earlier results
\cite{GZ1,GZ2,GZ3,GZS}.

Although in principle one can proceed further and under certain
approximations evaluate the matrix elements (\ref{me}) for pairs of time-reversed
paths, we will not do it here. The reason for that is obvious from the
above discussion: Particular values of $A_{1,2}^{nm}$ are irrelevant for
dephasing. It suffices to observe that these matrix elements do not grow
at long times.

\subsection{Relation to other results}

It is useful to compare eqs. (\ref{sigmafinal})-(\ref{me})
with the results of some earlier calculations of the WL correction.
Altshuler, Aronov and Khmelnitskii (AAK) \cite{AAK} considered the electron
dephasing by a fluctuating external field. Their results are easily recovered
from our calculation if one sets $V^- \equiv 0$. Then one again arrives
at eqs. (\ref{sigmafinal})-(\ref{me}) with
$A_{1,2}^{nm}\equiv\delta (y_{1,2}-z_{1,2})$.
AAK furthermore applied their results to the problem of
interacting electrons identifying an external field with one produced
by fluctuating electrons (the field $V^+$ in our analysis). In order
to account for Pauli blocking AAK suggested a phenomenological
procedure which amounts to keeping only the classical
part of this field and to cutting out its quantum modes with frequencies
$\omega >T$ (i.e. all modes at $T \to 0$). This last step has no
analogy in our calculation.

An attempt to justify the procedure
\cite{AAK} was recently undertaken by AAG \cite{AAG} within the
framework of the first order perturbation theory in the interaction.
Our general expressions, if expanded to the first order, yield
eq. (\ref{sigmapert2}) which can be written in the form
\begin{equation}
\sigma = \sigma^{(0)}+\delta\sigma^{(1)}_{\rm I}+ \delta\sigma^{(1)}_{\rm R}
\label{spert}
\end{equation}
Here $\sigma^{(0)}$ is the non-interacting contribution defined
by the first term in (\ref{sigmapert2}), $\delta\sigma^{(1)}_{\rm
I}$ corresponds to the terms in (\ref{sigmapert2}) which contain
the product $\hat V^+\hat V^+$ ($S_{\rm I}$-terms), while
$\delta\sigma^{(1)}_{\rm R}$ is given by the terms containing
$\hat V^+(1-2\hat \rho_0)\hat V^-$ and $(1-2\hat \rho_0)\hat V^-\hat
V^+$ ($S_{\rm R}$-terms).
As we have already shown in Ref. \cite{GZ3}, eq.  (\ref{sigmapert2})
is exactly equivalent to one derived by AAG \cite{AAG}.
In particular, it contains the combination
$\coth\frac{\hbar\omega}{2T}+\tanh\frac{\epsilon-\hbar\omega}{2T}$
leading to partial cancellation of the terms $\delta\sigma^{(1)}_{\rm I}$
and $\delta\sigma^{(1)}_{\rm R}$.

The perturbative result (\ref{sigmapert2}) is
reproduced at every stage of our analysis.
\begin{itemize}
\item{In order to obtain (\ref{sigmapert2}) one
can just evaluate the operators $\hat U_{1,2}$
perturbatively starting directly from their definition
(\ref{U12}). Substituting the result in eq. (\ref{sigma1})
and replacing $\hat\rho_V(t')\to\hat\rho_0$
in (\ref{sigma1}) one arrives at (\ref{sigmapert2}).}

\item{Alternatively, one can also expand the exact expressions
for $\hat U_{1,2}$ (\ref{exact}) in $V^\pm$. One recovers the same
result (\ref{sigmapert2}).}

\item{One can also expand approximate expressions for $\hat U_{1,2}$
(\ref{rhoVapp}) with the same result.}

\item{Finally, one can perform a perturbative expansion of the
quasiclassical result (\ref{sigmafinal}). One should expand $J^{nm}$
to the first order in $S_{\rm I}$ and $A^{nm}_{1,2}$ to the first
order in $V_0$. One obtains terms proportional to the functions
$\langle V^+V^+\rangle \to I$ (\ref{++})
and $\langle V^+V^-\rangle \to R$ (\ref{+-}). The structure of this first
order quasiclassical result is
identical to that of eq. (\ref{sigmapert2}), in the latter one
should just use the quasiclassical form (\ref{vV}) for the matrix
elements of the operators $\hat u_{1,2}$ and replace the coordinates in the
arguments of the fields $V^+$ by the classical
paths, $V^+(\tau_j,x_j)\to V^+(\tau_j,\tilde x_{1n,2m}(\tau_j))$.
Note that this substitution should be performed neither for the field $V^-$
nor for the electron density matrix $\hat \rho_0$ because no
quasiclassical approximation was employed with this matrix.
Further details are presented in Appendix.}
\end{itemize}

\subsection{Pauli principle and dephasing in the ground state}
It is sometimes argued that
electron decoherence at $T=0$ is impossible because of
the Pauli principle: Electron at the Fermi surface can
neither lose nor gain energy, hence, it cannot decohere.
Cancellation of ``coth'' and ``tanh'' terms in the
first order perturbation theory is considered by some authors
as a formal consequence of this energy constraint and, on the contrary,
independence of $\tau_{\varphi}$ on ``tanh'' terms is
interpreted as a sign of physical inconsistency
of the calculation (``Pauli principle is lost by approximations'').

Our analysis -- which fully accounts for the Pauli principle --
does not support the above point of view. Our final result,
eqs. (\ref{sigmafinal})-(\ref{me}), does depend on the Fermi function,
however, this dependence enters only into the pre-exponent via
the matrix elements $A_{1,2}^{nm}$ (\ref{me}) which in turn depend
on the electron density matrix $\hat \rho_0$. Thus,
the Pauli principle does not have any significant impact on the
dephasing process. As we have already explained
in Ref. \cite{GZ2} and elsewhere, electron dephasing
at low $T$ is only caused by fluctuations of the field $V^+$.
Such fluctuations are described by the $S_{\rm I}$-term in the
effective action which is not sensitive to $\rho_V$ at all. In the
presence of interactions the electron energy fluctuates and it remains
conserved only on average. At the same time electrons cannot, of course,
infinitely decrease their energies. Within our formalism such
process is prevented by the dissipative terms which explicitly depend
on $\rho_V$.
For instance, eq. (99) of Ref. \cite{GZ2} demonstrates that electrons
above the Fermi level decrease energies, however, for energies
below $\mu$ effective ``damping'' produced by the electron bath becomes
negative, and the holes are pushed up to the Fermi surface.
Such processes give rise to the time dependence of the pre-exponent
contained in the matrix elements $A_{1,2}^{nm}$ (\ref{me}).

In the arguments against quantum dephasing
at $T=0$ the Pauli principle is used merely as an energy constraint.
Therefore such arguments are not specific to Fermi systems
\cite{FN5} and can be tested for any quantum particle interacting with
a dissipative quantum environment. It is only important to ensure
that the whole interacting system ``particle+environment'' is in its
true ground state at $T=0$. One possible way to conduct such a test is
to study the equilibrium effect of persistent currents (PC) for a particle on a
ring in the presence of interactions. Since nonvanishing
PC can only exist in the presence of quantum coherence, (partial)
suppression of its amplitude by interactions may signal quantum dephasing.

Such a problem has recently been investigated by various authors and
suppression of PC by (long range) interactions was demonstrated
even at $T=0$ \cite{Buttiker,Paco,GHZ}
(see also \cite{GZ98}). In particular, for the model of a diffusive electron
gas \cite{Paco,GHZ} one finds that PC gets suppressed by interactions
exactly in the ground state provided the ring perimeter exceeds
a finite dephasing length $L_{\varphi}$ \label{GHZ}. This length turns out
to be fully consistent with one found from our WL analysis.

Without going into further details let us briefly address only one point
directly related to our discussion.
A non-perturbative instanton analysis of the problem \cite{GHZ} demonstrates
that suppression of PC by interactions is controlled by the parameter
\begin{equation}
\lambda \sum_{k=1}^rka_k \sim \lambda  r .
\label{param}
\end{equation}
Here $\lambda =3/(8k_F^2l^2) \ll 1$ is the dimensionless interaction
strength, $a_k$ are the Fourier coefficients of the interaction kernel
($a_k \sim (2/\pi r)\ln (r/k)$ for $1\leq k \lesssim r$ and
$a_k \approx 0$ otherwise) and $r=R/l \gg 1$ with $R$ being the ring radius.
Provided the
parameter (\ref{param}) is large, PC is strongly suppressed even at $T=0$.
The dephasing length  $L_{\varphi}$ is derived from the condition
$\lambda  r \sim 1$ which yields \cite{GHZ}  $L_{\varphi} \sim l/\lambda$.

Up to a numerical prefactor the parameter (\ref{param}) is just
the instanton action describing tunneling between two
different topological sectors of the problem. The
result (\ref{param}) cannot be correctly reproduced within
the perturbation theory. Indeed, let us expand the flux-depending part of the
free energy and PC to the first order in $\lambda$. 
Then at $T=0$ PC is found to be proportional
to the following combination \cite{GHZ}
\begin{equation}
\phi_x-\frac{\lambda}{2}\sum_{k=1}^rka_k
\ln \left(\frac{k+2\phi_x}{k-2\phi_x}\right),
\label{pert0}
\end{equation}
where $-0.5 <\phi_x \leq 0.5$ is the external flux (normalized to the flux
quantum) piercing the ring. For small $\phi_x \ll 1$ one can expand the logarithm and reduce
(\ref{pert0}) to
\begin{equation}
\phi_x\left[1-2\lambda\sum_{k=1}^r\frac{ka_k}{k}\right].
\label{pert1}
\end{equation}
The factors $k$ in the numerator and denominator cancel and
one is left only with a small correction
$2\lambda \sum_{k=1}^ra_k \sim \lambda  \ll 1$. The above cancellation
in eq. (\ref{pert1}) obtained within the Matsubara technique is to much extent
analogous to ``coth-tanh'' cancellation in the real time approach. In both
cases this cancellation is not complete, but the remaining term is small
and does not give the correct answer which can only be obtained by
non-perturbative means.

The above example provides yet one more illustration of insufficiency
of AAV's perturbative approach. For instance, following AAV's logics
one could qualify (\ref{param}) as ``an incorrect perturbative rather than
a nonperturbative'' result only because this parameter is proportional
to the first power of $\lambda$ but does not agree with one derived
from the perturbation theory (\ref{pert0}). Proceeding further
along these lines, one could also ``highlight'' a  ``mistake''
in our non-perturbative analysis \cite{GHZ}. In order to do so, one
could observe
that the same combination $\lambda \sum_{k=1}^rka_k$ enters
into both non-perturbative (\ref{param}) and perturbative (\ref{pert0})
expressions, however the latter also contains the logarithm which is
missing in the former. Following the logics of Ref. \cite{AAV} one would
then be led to conclude that the logarithm ``is omitted in all orders of
perturbation theory'' and the result (\ref{param}) ``is equivalent to
calculating only a single contribution'' $\lambda \sum_{k=1}^rka_k$.
In this way AAV arrived at their conclusion about
missing diagrams $\Sigma^{({\rm c-f})}$ in our calculation.

Fortunately, a detailed Monte Carlo simulation provides
a complete numerical solution for the problem \cite{GHZ}. It
unambiguously rules out the perturbative result (\ref{pert0}) and
demonstrates that PC is indeed strongly suppressed for $\lambda r \gg 1$
even exactly at $T=0$, see figs. 1 and 2 of Ref. \cite{GHZ}.
Similarly, our present results, eqs.
(\ref{sigmafinal})-(\ref{me}), allow to discard perturbative calculations
of the WL correction to conductivity at low temperatures.

\section{Remarks on AAV's critique}

The analysis of the previous section not only
rules out the AAV's claim about vanishing dephasing rate
at $T \to 0$ but also demonstrates that their critique of our calculation
\cite{GZ2,GZ3} is irrelevant. Nevertheless, for the sake of completeness
we will reply to both critical points ($i$) and $(ii)$ of Ref. \cite{AAV}.

\subsection{Density matrix}

In Ref. \cite{AAV} AAV stated that in eqs. (43) of our paper \cite{GZ2}
we ``replace the density matrix by its ``Wigner transform'''', eq. (5)
of Ref. \cite{AAV}. This AAV's statement is not correct. The only
replacement performed in eqs. (43) of \cite{GZ2} as compared to the exact
eqs. (40) of that paper (or eqs. (\ref{H12}) of this paper) is defined by
our present eq. (\ref{R0}), where
\begin{equation}
\hat\rho_0(\hat{\bm{p}},\hat{\bm{r}})=n(\hat H_0(\hat{\bm{p}},
\hat{\bm{r}}))
\label{rho}
\end{equation}
and $n (\xi)=1/[\exp (\xi/T)+1]$ is the Fermi function. In other words,
in Ref. \cite{GZ2} we used the following expressions
\begin{equation}
1-2\hat\rho_0(\hat{\bm{p}},\hat{\bm{r}})=
\tanh\left(\frac{\hat{\bm{p}}^2/2m-\mu +U_{\rm imp}(\hat{\bm{r}})}{2T}\right)
\label{rho5}
\end{equation}
and
\begin{equation}
\langle \bm{r}_1|1-2\hat\rho_0|\bm{r}_2 \rangle=\sum\limits_\nu
\tanh\frac{\xi_\nu}{2T} \psi_\nu (\bm{r}_1) \psi^*_\nu (\bm{r}_2),
\label{1-2rho}
\end{equation}
where $\xi_\nu$ and $\psi_\nu (\bm{r})$ are the eigenvalues and the
eigenfunctions of the Hamiltonian $\hat H_0$.
Eq. (\ref{rho5}) was used in eq. (43)
of \cite{GZ2} and further while constructing the effective action. Eq.
(\ref{1-2rho}) was used in Section 4 and Appendix A of Ref. \cite{GZ3}
(cf. eq. (54) of that paper)
while performing the first order perturbative calculation of
the conductance.
With the aid of the form (\ref{1-2rho}) in Ref. \cite{GZ3} we have proven
(partial) cancellation of ``coth'' and ``tanh'' terms in the first
order at $T \to 0$ and reproduced the results \cite{AAG}.
Also the combination $1-2n (\bm{p},\bm{r})$ in eqs. (52), (54)
and (68) of \cite{GZ2} has nothing to do with the
``Wigner transform'' of the density matrix, but is simply equal
to $\tanh (H_0(\bm{p},\bm{r})/2T)$. This form yields purely
real $S_{\rm R}$ for all paths and $S_{\rm R}=0$ for any pair of
time-reversed classical paths.

Let us compare our eqs. (\ref{rho})-(\ref{1-2rho}) with eq. (5) of
Ref. \cite{AAV}. The latter equation \cite{AAV} defines an object,
$\rho_{1'4}$, which is neither an operator (cf. our eqs.
(\ref{rho},\ref{rho5})) nor the electron
density matrix  $\rho_0 (\bm{r},\bm{r}'))=\langle
\bm{r}_1|\hat\rho_0|\bm{r}_2\rangle$ (cf. our eq. (\ref{1-2rho})).
We conclude that eq. (5) of Ref. \cite{AAV} has nothing to do with 
our analysis. Since AAV's
claim of our ``major mistake'' and their subsequent critique are
based on their eq. (5), both this claim and critique can be proven
irrelevant already by a direct comparison with what was {\it actually} done
in our paper \cite{GZ2}.

\subsection{Effective action and commutation relations}
In Ref. \cite{AAV} AAV pointed out that
while constructing our effective action we disregarded the Poisson brackets or,
which is the same, the commutation relations between the operators
$\hat \rho_0$ and $\hat V^-$
entering the Hamiltonians (\ref{H12}). AAV furthermore argued that
if one takes care about ordering of these operators, one arrives at the
effective action different from ours. Although the form of this action
was not specified, it was claimed in Ref. \cite{AAV} that the term $S_{\rm R}$
is not anymore real, but contains an imaginary part. According to AAV
this imaginary part provides nonzero contribution to $S_{\rm R}$
evaluated on pairs of time reversed paths and
``in perturbation theory ensures that the ultraviolet
divergence in $iS_R$ cancels that of $S_I$''.

The latter statement of AAV is false. The correct one
is just the opposite:
It is the {\it real} part of $S_{\rm R}$ that gives rise to ``tanh''-terms
which compensate ``coth''-contributions in the first order
perturbation theory
at $T=0$. AAV seem not to appreciate the fact that the term $iS_{\rm R}$
in the exponent of the influence functional and the matrix elements
generated by this term in the perturbation
theory are different mathematical objects. The perturbative contribution from a
purely {\it imaginary} term
$iS_{\rm R}$ can and does cancel the contribution from a purely {\it real}
term $S_{\rm I}$ in the first order at $T \to 0$ within the Golden rule
approximation. This is a general property not specific to any particular
calculation. For more information we refer to the textbook \cite{FH}
where the derivation of the perturbation theory from the
influence functional was analyzed in details, see eqs. (12-104) to (12-108)
of that book.

As to the commutation relations, everything is, of course, correct
with them in our path integral analysis. In order to demonstrate that
one should only keep track of correct ordering for the operators in
the perturbation expansion. One way could be to proceed directly with the
Hamiltonians (\ref{H12}) where ordering is defined uniquely and no
ambiguity can occur. Alternatively, the full perturbation
theory can be recovered by expanding
the influence functional in powers of $iS_{\rm R}+S_{\rm I}$. In this
case one should (a) replace the momentum and coordinate variables by
the operators $p \to \hat{p}$,
$r \to \hat{r}$ and (b) specify the proper way of
ordering (fixed by eqs. (\ref{H12})) {\it in addition} to the
expression for the effective action \cite{GZc}.

Furthermore, as it was demonstrated above, in order to
find $\tau_{\varphi}$ at $T \to 0$ it is sufficient to
correctly derive the classical ($\hbar$-independent) part of
the action only. Obviously, while deriving $S^{\rm (cl)}$
there is no need to take care of the commutation relations at all.
This action is always real and is obtained from the quantum Hamiltonian
by replacing the operators by the corresponding $c$-number functions.
For instance, from (\ref{H12})
(after the replacement $\hat\rho_V(t)\to\hat \rho_0$) we obtain
\begin{eqnarray}
S^{\rm (cl)}_{1,2}&=&\int_{t'}^{t} d\tau\left[
p\dot x-\frac{p^2}{2m}-U_{\rm imp}(x)+eV^+(\tau,x) \right.
\nonumber\\
&&\left.
\pm\,\frac{1}{2}[1-2n(H_0(p,x))]eV^-(\tau,x)\right].
\end{eqnarray}
These actions are real and insensitive to the ordering
of $e\hat V^-$ and $1-2\hat\rho_0$.
Hence, the action $S_{\rm R}$ \cite{GZ2} obtained from $S^{\rm (cl)}_{1,2}$
by averaging over $V^\pm,$ is real as well. On top of that, $S_{\rm R}$
vanishes on pairs of time-reversed paths. Hence,
it can only contribute to the pre-exponent.  The latter represents
the quantum correction which is sensitive to the ordering of the operators.
However, this correction is formally smaller in the parameter $\hbar$,
and, as we have already discussed in Sec. 4, it can never cancel
$S^{\rm (cl)}$ as long as the latter exceeds $\hbar$.

In our problem it is not convenient to apply the
van Vleck formula (\ref{vV}) directly to the Hamiltonians
$H_{1,2}$ (\ref{H12}). This is because the latter contain the
sharp function of the electron momentum $1-2n(H_0(p,x))$ which
effectively turns fluctuations around the classical paths non-Gaussian.
This -- purely technical -- complication is circumvented by eqs.
(\ref{exact}) and the subsequent analysis of Sec. 5. This analysis
demonstrates that non-Gaussian fluctuations give rise to the
pre-exponential factors
$A_{1,2}$ in the expression for the conductivity (\ref{sigmafinal}).
We have proven in Sec. 5 that these factors are irrelevant for
dephasing because they do not grow at long times and, hence, cannot
cancel the term $S_{\rm I}$.

\section{Summary}

In summary, we have derived a complete expression for the weak
localization correction to
the conductivity of a disordered conductor in the presence
electron-electron interactions. Our analysis has been carried out
within the standard model for an interacting electron gas in disordered
conductors with no approximations beyond the accuracy of the
definition of the WL correction. In particular, interactions have
been treated non-perturbatively, no quasiclassical
approximation for the electron density matrix has been employed
and no disorder averaging has been performed at all. We have fully
confirmed our earlier results \cite{GZ1,GZ2,GZ3,GZS} and extended them by
explicitly taking into account quantum
fluctuations around the classical paths for interacting electrons.
We have proven that such fluctuations, while practically
irrelevant for the calculation of $\tau_{\varphi}$, do contribute
to the Cooperon dynamics at short times causing, for instance,
partial cancellation of the well known ``coth'' and ``tanh'' terms
in the first order perturbation theory. We have also demonstrated
the failure of a perturbative calculation \cite{AAV} in the problem of
quantum dephasing of electrons at low temperatures. Finally we have
refuted AAV's critique of our previous calculation \cite{GZ2,GZ3} observing
that ($i$) Poisson brackets are irrelevant for the problem of
electron dephasing by interactions and ($ii$) no ``Wigner transform''
of the electron density matrix was performed in our derivation.

\section{Acknowledgment}

We are grateful to G. Sch\"on for numerous instructive discussions.
This work is part of the Kompetenznetz ``Funktionelle Nanostructuren''
supported by the Landestiftung Baden-W\"urttemberg gGmbH.

{\it Note added.} After this paper had already been submitted there
appeared an independent work \cite{vD} addressing the same issue.
In this work von Delft (vD) has successfully re-derived our influence 
functional for interacting electrons \cite{GZ2} and argued that
(a) our approach ``properly incorporates the Pauli principle'' and
(b) ``the standard Keldysh diagrammatic expressions for the self
energy of the Cooperon can be obtained from $iS_{\rm R}+S_{\rm I}$'',
i.e. from our influence functional. Thus, it is now verified
not only in our work but also independently by other authors
\cite{vD,Er} that our path integral result (\ref{J}) \cite{GZ2} contains
{\it all} RPA diagrams to {\it all} orders in the electron-electron
interaction. 

The observations (a) and (b) are important because they 
allow to restrict the whole discussion to just one -- purely
mathematical -- issue, i.e. how to correctly evaluate the path integral
(\ref{J}). We believe that the analysis 
presented in Sec. 5 of this paper should eliminate all doubts
\cite{vD} concerning the role of $S_{\rm R}$-terms for quantum 
dephasing of electrons at low temperatures. This analysis, for instance, 
involves {\it none} of the approximations denoted in \cite{vD} as (i), (ii) and
(iii). In particular, it rules out vD's conjecture that within our
approach we ``neglect all the diagrams of Fig. 2b'' of
\cite{vD}. Quite on the contrary, our final result,
eqs. (\ref{sigmafinal})-(\ref{me}), explicitly accounts for all these 
diagrams (giving rise to the ``tanh''-contribution
(\ref{deltasigma2}) in the first order) as well as for 
infinitely many diagrams of all higher orders not presented in Ref. \cite{vD}.

Several additional comments are in order: ($i$) The statement
\cite{vD} that the first order perturbative result contains no
ultraviolet (UV) divergences is explicitly incorrect for 
1d and 2d systems, see, e.g., eq. (70) of Ref. \cite{GZ3}.
($ii$) We disagree with vD's conjecture 
(see the footnote 16 in Ref. \cite{vD})
that diagrams with crossed and overlapping interaction lines can
be neglected \cite{FN6}.
($iii$) Further evidence that the above conjecture is problematic is
provided by the results \cite{GHZ} which we also address in
Sec. 5E \cite{FNvD}. In that problem the first order diagrams
yield negligible contribution (\ref{pert1}) and, hence,
the correct result (\ref{param}) is dominated by {\it all} the 
remaining diagrams with crossed and overlapping interaction lines. 
($iv$) The argument presented by eqs. (10-11) of \cite{vD} is
inconclusive, since it is based on an improper application of
quasiclassical methods to Fermi systems \cite{FN9}. If one expands 
the electron density matrix $\rho_0$ in powers of $\hbar$ one indeed gets
a series of terms diverging at $T \to 0$. However, this observation
can only imply
that the Taylor expansion of the step function (i.e. the Fermi function at 
$T \to 0$ and energies close to $\mu$) is essentially useless.  
A much more useful strategy 
is to retain $\rho_0$ (which is, of course, always finite and not
large) in its full quantum mechanical form and to apply quasiclassics 
only to those matrix elements which do not contain $\rho_0$. 
This strategy was implemented in Sec. 5.
($v$) In contrast to vD's conjecture in the footnote 23 of \cite{vD}
our result (\ref{sigmafinal})-(\ref{me}) does {\it not} diverge
at $T \to 0$.

\appendix

\section{}
Here we will summarize several expressions used in our
calculation and present some perturbative in the interaction
results for the conductivity.

Averaging over the fluctuating fields $V^{\pm}$ implies calculating
the double path integral
\begin{equation}
\langle \dots \rangle_V=\int {\cal D}V^+ {\cal D}V^-(...)
\exp\left\{ \frac{i}{\hbar}S_{EM}[V^\pm]\right\} .
\label{av}
\end{equation}
Here $S_{EM}$ can be understood as a formally exact effective action
for the Hubbard-Stratonovich fields $V^{\pm}$, see, e.g., eq. (10)
of Ref. \cite{GZ2}. For the situation discussed here it is sufficient
to expand $S_{EM}$ to the second order in $V$. Then one finds
\begin{eqnarray}
S_{EM}[V^\pm]=\int\frac{d^4K}{(2\pi)^4}V^-(-K)\frac{k^2\epsilon(K)}{4\pi}V^+(K)
\hspace{0.7cm}
\nonumber\\
+\frac{i}{2}\int\frac{d^4K}{(2\pi)^4}V^-(-K)\frac{k^2{\rm Im}\epsilon(K)}{4\pi}
\coth\frac{\hbar\omega}{2T}V^-(K),
\label{Sem}
\end{eqnarray}
where $K=(\omega,k).$
This action allows to determine the correlation functions
\begin{eqnarray}
\langle V^+(t,{r})V^+(0,0)\rangle=\hbar
I(t,{r})\hspace{3.3cm}
\nonumber\\
=\hbar\int\frac{ d^4K}{(2\pi)^4}\,
{\rm Im}\left(\frac{-4\pi}{k^2\epsilon(K)} \right)\coth\frac{\hbar\omega}{2T}\,e^{-iKX},
\label{++}\\
\langle V^+(t,r)V^-(0,0)\rangle=i\hbar 
R(t,{r})\hspace{3cm}
\nonumber\\
=i\hbar\int\frac{ d^4K}{(2\pi)^4}\,
\frac{4\pi}{k^2\epsilon(K)} \,e^{-iKX},
\label{+-}\\
\langle V^-(t,{r})V^-(0,0)\rangle= 0. \hspace{4.1cm}
\end{eqnarray}
Here we have defined $X=(t,r),$ and $KX=\omega t-kr.$
Employing the action (\ref{Sem}) is equivalent to describing the
electron-electron interaction within the random phase approximation.

Let us expand the results (\ref{sigma1},\ref{exact}) to the
first non-vanishing order in the interaction (second order in $V^\pm$)
and replace $\hat\rho_V(t')=\hat\rho_0$. Then we obtain 
\begin{eqnarray}
\sigma =-\frac{ie}{3\hbar}\int\limits_{-\infty}^t dt'
\left\langle{\rm tr}\left(\hat{  {j}}({x})
\hat u_0(t,t')
[\hat{  {x}},\hat \rho_0]
\hat u_0(t',t)
\right)\right\rangle_{V}
\nonumber\\
-\,\frac{2e^3}{3\hbar^3}\,\int\limits_{-\infty}^t dt'
\int\limits_{t'}^t d\tau_1\int\limits_{t'}^{\tau_1} d\tau_2
{\rm Im}\bigg[
\bigg\langle{\rm tr}\bigg(\hat{  {j}}({x})
\hat u_0(t,\tau_1)\hspace{0.6cm}
\nonumber\\
\times \hat V^+(\tau_1)\hat u_0(\tau_1,\tau_2)
\left\{\hat V^+(\tau_2) +\frac{1}{2}(1-2\hat\rho_0)\hat V^-(\tau_2) \right\}
\nonumber\\
\times\hat u_0(\tau_2,t')
[\hat{  {x}},\hat \rho_0]
\hat u_0(t',t)
\bigg)
\nonumber\\
+
{\rm tr}\bigg(\hat{{j}}({x}) 
\hat u_0(t,\tau_2)
\left\{\hat V^+(\tau_2)+\frac{1}{2}(1-2\hat\rho_0)\hat V^-(\tau_2)\right\}
\nonumber\\
\hat u_0(\tau_2,t')
[\hat{{x}},\hat \rho_0]
\hat u_0(t',\tau_1)\hat V^+(\tau_1)\hat u_0(\tau_1,t)
\bigg)\bigg\rangle_{V}
\bigg].
\label{sigmapert2}
\end{eqnarray}
After averaging over $V^\pm$ the latter expression coincides with the result
derived in Ref. \cite{AAG}. The term proportional to 
$\coth\frac{\hbar\omega}{2T}$ emerges from
the average $\langle\hat V^+(\tau_1)\hat V^+(\tau_2)\rangle$,
while  $\tanh\frac{\xi-\hbar\omega}{2T}$ appears from
the combination $1-2\hat\rho_0$. Further details can be found in 
Ref. \cite{GZ3}.

Note, that eq. (\ref{sigmapert2}) represents the exact first order result
obtained without any evaluation of the path integrals. Eq.
(\ref{sigmafinal}) is valid to all orders in the interaction, but it 
was derived by evaluating the path integrals
in the quasiclassical limit $k_Fl \gg 1$. Let us expand $J^{nm}$ (\ref{Jevol})
to the first order in $S_{\rm I}$ and $A^{nm}_{1,2}$ (\ref{me}) to the first
order in $V_0$. Then with the aid of eq. (\ref{sigmafinal}) we reproduce 
eq. (\ref{sigmapert2}) with trivial modifications as described towards
the end of Sec. 5D. 

Let us now identically transform our first order quasiclassical results for 
$\delta\sigma_{\rm I,R}^{(1)}$ to a somewhat different form. For that
purpose let us express the electron density matrix $\rho_0$ as follows:
\begin{eqnarray}
\langle z_1|\hat\rho_0|z_2\rangle=\rho_0(z_1,z_2)
\hspace{4.35cm}
\nonumber\\
=\frac{1}{2}\int\limits_{-\infty}^{+\infty}ds_2
\left[\delta(s_2)+\frac{iT}{\hbar\sinh[\frac{\pi Ts_2}{\hbar}]}\right]
u_0(-s_2,z_1,z_2),
\label{transform}
\end{eqnarray}
where $u_0(-s,z_1,z_2)=\langle z_1|\hat u_0(0,s)|z_2\rangle$
is the matrix element of the evolution operator for
non-interacting electrons.
In addition we note that the functions $I(t,r)$ (\ref{++}) and
$R(t,r)$ (\ref{+-}) are related to each other by means of the identity
$$
I(X)=\int
ds_1 \frac{T\coth\frac{\pi Ts_1}{\hbar}}{2\hbar}
[R(t-s_1,r)+R(-t-s_1,r)].
$$
Making use of the above identities we arrive at the following expressions for
$\delta\sigma_{\rm I,R}^{(1)}$:
\begin{widetext}
\begin{eqnarray}
\delta\sigma_{\rm I}^{(1)}&=&\frac{e^4}{12 m\hbar}
\sum_{n,m}\int_{-\infty}^tdt'\int_{-\infty}^{+\infty}ds_1ds_2\,\int dy_1 dy_2\,
\frac{-iT^2\coth[\pi Ts_1/\hbar]}{\hbar^2\sinh[{\pi Ts_2}/{\hbar}]}\,(\nabla_{x_1}-\nabla_{x_2})|_{x_1=x_2}
\nonumber\\ &&\times\,
\frac{1}{(2\pi\hbar)^3}
\sqrt{{\rm det}\frac{\partial^2 S^{(n)}_0(t,t',x_1,y_1)}{\partial x_{1}\partial y_{1}}\,
{\rm det}\frac{\partial^2 S^{(m)}_0(t,t',x_{2},y_{2})}{\partial x_{2}\partial y_{2}}}
\nonumber\\ &&\times\,
\exp\left\{\frac{i}{\hbar}S^{(n)}_0(t,t',x_1,y_1)-\frac{i}{\hbar}S^{(m)}_{0}(t,t',x_2,y_2)
\right\}
\nonumber\\ && \times\,
\bigg\{ \int_{t'}^td\tau_1\int_{t'}^{t}d\tau_2
\bigg[ R(\tau_1-\tau_2-s_1,\tilde x_{1n}(\tau_1)-\tilde x_{1n}(\tau_2))+ R(\tau_1-\tau_2-s_1,\tilde x_{2m}(\tau_1)-\tilde x_{2m}(\tau_2))
\nonumber\\ &&
-\,R(\tau_1-\tau_2-s_1,\tilde x_{1n}(\tau_1)- \tilde x_{2m}(\tau_2))-R(\tau_1-\tau_2-s_1,\tilde x_{2m}(\tau_1)-\tilde x_{1n}(\tau_2))\bigg] \bigg\}
\nonumber\\ &&\times\,
(y_1-y_2) u_0(-s_2,y_1,y_2)
\label{deltasigma1}
\end{eqnarray}
and
\begin{eqnarray}
\delta\sigma_{\rm R}^{(1)}&=&\frac{e^4}{12 m\hbar}
\sum_{n,m}\int_{-\infty}^tdt'\int_{-\infty}^{\infty} ds_1ds_2\int dy_1 dy_2 dz dr\,
\frac{T}{\hbar\sinh[{\pi Ts_1}/{\hbar}]}\frac{-iT}{\hbar\sinh[{\pi Ts_2}/{\hbar}]}\,
\nonumber\\ &&\times\,
(\nabla_{x_1}-\nabla_{x_2})|_{x_1=x_2}\frac{1}{(2\pi\hbar)^3}
\sqrt{{\rm det}\frac{\partial^2 S^{(n)}_0(t,t',x_1,y_1)}{\partial x_{1}\partial y_{1}}\,
{\rm det}\frac{\partial^2 S^{(m)}_0(t,t',x_{2},y_{2})}{\partial x_{2}\partial y_{2}}}
\nonumber\\ && \times\,
\exp\left\{\frac{i}{\hbar}S^{(n)}_0(t,t',x_1,y_1)-\frac{i}{\hbar}S^{(m)}_{0}(t,t',x_2,y_2)
\right\}
\nonumber\\ && \times\,
\bigg\{\int_{t'}^td\tau_1 \int_{t'}^t d\tau_2\,\bigg[
u_0(t'-\tau_2-s_1,y_1,r)\big[R(\tau_1-\tau_2,\tilde x_{1n}(\tau_1)-r)
\nonumber\\&&
-\,R(\tau_1-\tau_2,\tilde x_{2m}(\tau_1)-r)\big]u_0(\tau_2-t',r,z)(z-y_2)u_0(-s_2,z,y_2)
\nonumber\\
&&-\,
(y_1-z) u_0(-s_2,y_1,z)u_0(t'-\tau_2,z,r)
\big[R(\tau_1-\tau_2,\tilde x_{1n}(\tau_1)-r)
\nonumber\\&&
-\,R(\tau_1-\tau_2,\tilde x_{2m}(\tau_1)-r)\big]
u_0(\tau_2-t'-s_1,r,y_2)\bigg]\bigg\}.
\label{deltasigma2}
\end{eqnarray}
\end{widetext}
Eqs. (\ref{deltasigma1}) and
(\ref{deltasigma2}) have a very similar structure. These
two expressions are, however, not fully identical even at $T \to 0$
and, hence, they {\it do not} cancel exactly in
the first order result (\ref{spert}), see also Sec. 4 of Ref. \cite{GZ3}.

Further evaluation of eqs.  (\ref{deltasigma1}) and (\ref{deltasigma2})
makes little sense because the first order perturbation theory cannot
provide any useful information about the electron dephasing time at
low temperatures. Nevertheless, the above expressions are of a certain
interest, since they help to illustrate the relation between perturbative
and non-perturbative results at the stage when the quasiclassical approximation
has already been performed. We observe, for instance, that all the first order
terms, both ``coth'' and ``tanh'' contributions, are fully reproduced from our
path integral analysis. Another observation concerns the relation between
the quasiclassical paths emerging from the path integrals and those entering
the first order results for $\delta\sigma_{\rm R}^{(1)}$. The WL correction
to the conductivity is defined on pairs of time-reversed path, and only such
paths (plus fluctuations around them) are relevant for the path integral
analysis of this quantity. Of course, the same paths enter if the general
result is expanded to the first order in the interaction before the
transformation (\ref{transform}). However, {\it after} this transformation
there appear additional matrix elements $u_0$ of the electron evolution
operator. Proceeding quasiclassically, one can evaluate these matrix
elements by means of the van Vleck formula (\ref{vV}), i.e. to write
\begin{eqnarray}
u_0(t'-\tau_2-s_1,y_1,r)
\propto e^{\frac{i}{\hbar}S^{(k)}_0(t'-\tau_2-s_1,0,y_1,r)}
\label{ill}
\end{eqnarray}
and similarly for other matrix elements. Substituting $u_0$ in the
form (\ref{ill}) into eq. (\ref{deltasigma2}) one
can interpret the result in terms of the electron motion along
additional classical paths $\tilde x_k(s)$, say, first from
$y_1$ to $r$ and then from $r$ to $z$ (some of these paths violate the
requirement of
causality, see Fig. 3 of Ref.
\cite{GZ3} and related discussion). This could in turn create an illusion
that these additional paths are missing in the path integral formulation.
The above analysis clearly indicates the origin of such an illusion. It also
demonstrates that -- in contrast to Ref. \cite{AAG} -- the whole
issue has nothing to do with disorder averaging which is not performed here
at all.




\begin{thebibliography}{99}
\bibitem{FN0} The third AAV's claim, that our path integral analysis does
not reproduce the perturbation theory \cite{AAG}, is fully based on
their false statements ($i$) and ($ii$) and, hence, is in error as well.
We also note that all the statements made by AAV in the footnote 
9 of Ref. \cite{AAV}
are incorrect and misleading. For instance, we never argued that ``Hikami boxes
appearing in Fig. 3'' of Ref. \cite{AAV} vanish. Rather, as one can easily
check in Sec. IV of Ref. \cite{GZS}, we argued that Hikami boxes do not
appear in our path integral approach at all.
\bibitem{sec} The second part of Ref. \cite{AAV} is devoted to an
alternative derivation of the first order perturbative results
\cite{AAG}. As such, this derivation can hardly be of any use for the
discussion of our nonperturbative analysis.
\bibitem{FN1} For the case of weak interactions considered here the
density matrix $\hat \rho_V$ is close to the Fermi function, 
cf. eqs. (\ref{rho}-\ref{1-2rho}). As this function tends to the step
function of energy at $T \to 0$, almost all the eigenvalues of $\hat
\rho$ are close to either one or zero.
\bibitem{FN2} For the sake of definiteness and simplicity
we use the substitution (\ref{R0}),
in which case averaging over the fields $V^\pm$ is performed exactly.
One can also avoid this step and -- as it was done in Sec. 5B -- prove
that averaging of the terms containing $\hat \rho_V(t')$ can be
performed separately from that of the matrix elements of $\hat u (t,t',V^+)$.
For $k_Fl \gg 1$ the accuracy of this procedure is the same as that of
eqs. (\ref{sigmafinal})-(\ref{me}).
\bibitem{FN5} For instance, ``qualitative arguments'' of Ref. \cite{AAG}
against quantum dephasing at $T \to 0$ are formulated for an arbitrary
quantum particle interacting with a bath of oscillators. These arguments are
not specific to Fermi statistics. For further discussion of this and
related issues see, e.g., Refs. \cite{Weiss,GZ98,Nag}.
\bibitem{FN6} Unfortunately, no calculation supporting this 
conjecture was presented in Ref. \cite{vD}. The only argument \cite{vD},
that such diagrams contain at least one extra power of the 
inverse dimensionless conductance $1/g$, cannot be accepted. 
First of all, it was not specified in \cite{vD} at 
which length scale the dimensionless conductance $g$ should be 
calculated (clearly, $1/g$ can become large, e.g., for sufficiently 
long wires). Since the parameter $1/g$ does not even exist before disorder 
averaging and since the diagrammatic expansion is carried out
in powers of the interaction, it would be more appropriate to say that the
diagrams in question contain an extra power of the effective interaction
parameter, but not $1/g$. More importantly, an extra power of a small parameter 
in front of the diagram does not yet imply that its contribution is small.
\bibitem{FNvD} We disagree with vD who argued that any
comparison between the models with and without the Pauli exclusion
principle would be ``meaningless'' \cite{vD}. Provided the interaction
term in the Hamiltonian is the same, the two models are very
similar in many respects, both physically and formally. 
For instance, the structure of all the diagrams is the same in all orders, 
and the zero temperature cancellation of (some) UV-divergent first order 
terms occurs in the 
same manner both with and without the Pauli principle  
(cf., e.g., eqs. (67) and (E19) of Ref. \cite{GZ3}).
\bibitem{FN9} Should this argument be correct, it would equally
apply to our theory and to that of AAK \cite{AAK}.
\end{thebibliography}
\end{document}